\documentclass[superscriptaddress,a4paper,12pt]{revtex4}
\usepackage{amsfonts}
\usepackage{amsmath}
\usepackage{amssymb}

\input{tcilatex}
\begin{document}

\author{Olimpia Lombardi}
\affiliation{CONICET-Universidad de Buenos Aires, Argentina}
\author{Juan Sebasti\'{a}n Ardenghi}
\affiliation{CONICET-IAFE-Universidad de Buenos Aires, Argentina}
\author{Sebastian Fortin}
\affiliation{CONICET-IAFE-Universidad de Buenos Aires, Argentina}
\author{Martin Narvaja}
\affiliation{CONICET-Universidad de Buenos Aires, Argentina}

\begin{abstract}
In this paper we review Castagnino's contributions to the foundations of
quantum mechanics. First, we recall his work on quantum decoherence in
closed systems, and the proposal of a general framework for decoherence from
which the phenomenon acquires a conceptually clear meaning. Then, we
introduce his contribution to the hard field of the interpretation of
quantum mechanics: the modal-Hamiltonian interpretation solves many of the
interpretive problems of the theory, and manifests its physical relevance in
its application to many traditional models of the practice of physics. In
the third part of this work we describe the ontological picture of the
quantum world that emerges from the modal-Hamiltonian interpretation,
stressing the philosophical step toward a deep understanding of the
reference of the theory.
\end{abstract}

\title{Foundations of quantum mechanics: decoherence and interpretation}
\maketitle

\section{Introduction}

Anybody who has been close to Prof. Mario Castagnino, even for a short time,
knows that he is an ever-eager spirit: the many different subjects treated
in this issue are a clear manifestation of the wide panoply of interests
that have moved him during his long academic life. Nevertheless, the present
article has a peculiarity with respect to the rest of the papers of the
issue: Castagnino should be one of the authors of this work. In fact, since
ten years ago he has been actively engaged with the foundations and the
philosophy of physics, leading an always increasing research group to which
we belong. In this field we have obtained relevant results with a remarkable
repercussion.

As Castagnino uses to say, he is a senior physicist but a baby philosopher.
However, this fact was not an obstacle to his eager spirit, which has been
involved in the foundations of so many different subjects that cannot be
addressed in a single article. In the present paper we will confine our
attention to Castagnino's contributions to the foundations of quantum
mechanics (QM), in order to review his main results in this area. First, we
will recall his work on quantum decoherence in closed systems, and the
proposal of a general framework for decoherence from which the phenomenon
acquires a conceptually clear meaning. Then, we will introduce his
contribution to the hard field of the interpretation of QM: our
modal-Hamiltonian interpretation solves many of the interpretive problems of
the theory, and manifests its physical relevance in its application to many
traditional models used in the practice of physics. In the third part of
this work we will describe the ontological picture of the quantum world that
emerges from our interpretation; here we will stress our philosophical step
toward a deep understanding of the reference of the theory, a move not usual
in the contemporary discussions about the interpretation of QM. Finally, we
will briefly recall Castagnino's contributions in other areas of the
foundations and the philosophy of physics.

\section{Foundations of quantum decoherence}

More than a decade ago Castagnino developed, with Roberto Laura, a formalism
that explains the limit reached by expectation values in closed quantum
systems with continuous spectrum,\cite{CL-1}$^{-}$\cite{CL-6} and begun to
conceive that formalism in terms of decoherence. When, some years later,
those works were reanalyzed in the context of our research group, we
acknowledged the conceptual relevance and the fruitful perspectives of that
work. So, the original proposal was further elaborated from a conceptual
viewpoint, and presented in several meetings and papers.\cite{SID-1}$^{-}$%
\cite{SID-10} In particular, we were invited by Prof. Fred Kronz, from the
University of Texas at Austin, to discuss that new view, and he suggested
the name `self-induced decoherence' (SID)\ in contrast with the orthodox
`environment-induced decoherence' (EID) approach.\cite{Paz-Zurek}$^{,}$\cite%
{Zurek-2003}

In those first works, we presented SID as different from EID, that is, as
the way in which decoherence manifests itself in closed systems. However,
shortly after we realized that both approaches can be subsumed under a 
\textit{General Theoretical Framework for Decoherence} (GTFD), which
encompasses decoherence in open and closed systems.\cite{GTFD-1}$^{-}$\cite%
{MPLA-random} According to this framework, decoherence is just a particular
case of the general phenomenon of irreversibility in QM.\cite{Omnes-2001}$%
^{,}$\cite{Omnes-2002} Since the quantum state $\rho (t)$ follows a unitary
evolution, it cannot reach a final equilibrium state for $t\rightarrow
\infty $. Therefore, if we want to explain the emergence of non-unitary
irreversible evolutions, we must split the whole space $\mathcal{O}$ of all
possible observables into a relevant subspace $\mathcal{O}_{R}\subset 
\mathcal{O}$ and an irrelevant subspace. With this strategy we restrict the
maximal information about the system: the expectation values $\langle
O_{R}\rangle _{\rho (t)}$ of the observables $O_{R}\in \mathcal{O}_{R}$
express that relevant information. Of course, the decision about which
observables are to be considered as relevant depends on the particular
purposes in each situation; but without this restriction, irreversible
evolutions cannot be described. In fact, the different approaches to
decoherence always select a set $\mathcal{O}_{R}$ of relevant observables in
terms of which the time behavior of the system is described: gross
observables in van Kampen,\cite{van Kampen} macroscopic observables of the
apparatus in Daneri \textit{et al.},\cite{Daneri} observables of the open
system in EID,\cite{Zeh-1}$^{-}$\cite{Zurek-2003} relevant observables in Omn%
\'{e}s.\cite{Omnes-1994}$^{,}$\cite{Omnes-1999}

Once the essential role played by the selection of the relevant observables
is clearly understood, decoherence can be explained in three general steps:

\begin{enumerate}
\item[1.] \textbf{First step:} The set $\mathcal{O}_{R}$ of relevant
observables is defined.

\item[2.] \textbf{Second step:} The expectation value $\langle O_{R}\rangle
_{\rho (t)}$, for any $O_{R}\in \mathcal{O}_{R}$, is obtained. This step can
be formulated in two different but equivalent ways:

\begin{itemize}
\item $\langle O_{R}\rangle _{\rho (t)}$ is computed as the expectation
value of $O_{R}$ in the unitarily evolving state $\rho (t)$.

\item A coarse-grained state $\rho _{G}(t)$ is defined by $\langle
O_{R}\rangle _{\rho (t)}=\langle O_{R}\rangle _{\rho _{G}(t)}$ for any $%
O_{R}\in \mathcal{O}_{R}$, and its non-unitary evolution (governed by a
master equation) is computed.
\end{itemize}

\item[3.] \textbf{Third step:} It is proved that $\langle O_{R}\rangle
_{\rho (t)}=\langle O_{R}\rangle _{\rho _{G}(t)}$ reaches a final
equilibrium value $\langle O_{R}\rangle _{\rho _{\ast }}$: 
\begin{equation}
\lim_{t\rightarrow \infty }\langle O_{R}\rangle _{\rho
(t)}=\lim_{t\rightarrow \infty }\langle O_{R}\rangle _{\rho _{G}(t)}=\langle
O_{R}\rangle _{\rho _{\ast }}  \label{2-1}
\end{equation}

where the final equilibrium state $\rho _{\ast }$ is obviously diagonal in
its own eigenbasis, which turns out to be the final pointer basis. But the
unitarily evolving quantum state $\rho (t)$ of the whole system \textit{has
only a} \textit{weak limit}: 
\begin{equation}
W-\lim_{t\rightarrow \infty }\rho (t)=\rho _{\ast }  \label{2-2}
\end{equation}

This weak limit means that, although the off-diagonal terms of $\rho (t)$
never vanish through the unitary evolution, the system decoheres \textit{%
from an observational point of view}, that is, from the viewpoint given by
any relevant observable $O_{R}\in \mathcal{O}_{R}$.
\end{enumerate}

This GTFD allows us to face the conceptual challenges that the EID approach
still has to face. One of them comes from the fact that, since the
environment may be \textquotedblleft external\textquotedblright\ or
\textquotedblleft internal\textquotedblright , the EID approach offers no
general criterion to decide where to place the \textquotedblleft
cut\textquotedblright\ between system and\ environment. Zurek considers this
fact as a shortcoming of his proposal: \textquotedblleft \textit{In
particular, one issue which has been often taken for granted is looming big,
as a foundation of the whole decoherence program. It is the question of what
are the `systems' which play such a crucial role in all the discussions of
the emergent classicality.}\textquotedblright \cite{Zurek-cut} In order to
address this problem, the first step is to realize that the EID relevant
observables of the closed system $U$ are those corresponding to the open
system $S$: 
\begin{equation}
O_{R}=O_{S}\otimes \mathbb{I}_{E}\in \mathcal{O}_{R}\subset \mathcal{O}
\label{2-3}
\end{equation}%
where $O_{S}\in \mathcal{O}_{S}$ of $S$ and $\mathbb{I}_{E}$ is the identity
operator in $\mathcal{O}_{E}$ of $E$. \ The reduced density operator $\rho
_{S}(t)$ of $S$ is defined by tracing over the environmental degrees of
freedom, 
\begin{equation}
\rho _{S}(t)=Tr_{E}\,\rho (t)  \label{2-4}
\end{equation}%
The EID approach studies the time-evolution of $\rho _{S}(t)$ governed by an
effective master equation; it proves that, under certain definite
conditions, $\rho _{S}(t)$ converges to a stable state $\rho _{S\ast }$: $%
\rho _{S}(t)\longrightarrow \rho _{S\ast }$. But we also know that the
expectation value of any $O_{R}\in \mathcal{O}_{R}$ in the state $\rho (t)$
of $U$ can be computed as 
\begin{equation}
\langle O_{R}\rangle _{\rho (t)}=Tr\,\left( \rho (t)(O_{S}\otimes \mathbb{I}%
_{E})\right) =Tr\left( \rho _{S}(t)\,O_{S}\right) =\langle O_{S}\rangle
_{\rho _{S}(t)}  \label{2-6}
\end{equation}%
Therefore, the convergence of $\rho _{S}(t)$ to $\rho _{S\ast }$ implies the
convergence of the expectation values: 
\begin{equation}
\langle O_{R}\rangle _{\rho (t)}=\langle O_{S}\rangle _{\rho
_{S}(t)}\longrightarrow \langle O_{S}\rangle _{\rho _{S\ast }}=\langle
O_{R}\rangle _{\rho _{\ast }}  \label{2-7}
\end{equation}%
where $\rho _{\ast }$ is a final diagonal state of $U$, such that $\rho
_{S\ast }=Tr_{E}\,\rho _{\ast }$.

>From this new conceptual perspective, we have studied the well-known
spin-bath model: a closed system $U=P\cup P_{1}\cup \ldots \cup P_{N}=P\cup
(\cup _{i=1}^{N}P_{i})$, where (i) $P$ is a spin-1/2 particle represented in
the Hilbert space $\mathcal{H}_{P}$, and (ii) each $P_{i}$ is a spin-1/2
particle represented in its Hilbert space $\mathcal{H}_{i}$. The Hilbert
space of the composite system $U$\ is, then, 
\begin{equation}
\mathcal{H}=\mathcal{H}_{P}\otimes \left( \bigotimes\limits_{i=1}^{N}%
\mathcal{H}_{i}\right)  \label{2-8}
\end{equation}%
If the self-Hamiltonians $H_{P}$ of $P$ and $H_{i}$ of $P_{i}$ are taken to
be zero, and there is no interaction among the $P_{i}$, then the total
Hamiltonian $H$ of the composite system $U$ is given by the interaction
between the particle $P$ and each particle $P_{i}$.\cite{Zurek-1982} By
contrast to the usual presentations, we have studied different
decompositions of the whole closed system $U$ into a relevant part and its
environment.\cite{MPLA}

\paragraph{Decomposition 1: A large environment that produces decoherence.}

In the typical situation studied by the EID approach, the open system $S$ is
the particle $P$, and the remaining particles $P_{i}$ play the role of the
environment $E$: $S=P$ and $E=\cup _{i=1}^{N}P_{i}$. This decomposition
results in the system decoherence when the number of particles in the bath
is very large.

\paragraph{Decomposition 2: A large environment with no decoherence}

We can conceive different ways of splitting the whole closed system $U$. For
instance, we can decide to observe a particular particle $P_{j}$ of what was
previously considered the environment, and to consider the remaining
particles as the new environment, in such a way that $S=P_{j}$ and $E=P\cup
(\cup _{i=1,i\neq j}^{N}P_{i})$. This decomposition results in that the
system does not decohere.

\paragraph{Decomposition 3: A small environment that produces decoherence}

It may be the case that the measuring arrangement \textquotedblleft
observes\textquotedblright\ a subset of the particles of the environment,
e.g., the $p$ first particles $P_{j}$. \ In this case, the system of
interest is composed by $p$ particles, $S=\cup _{i=1}^{p}P_{i}$, and the
environment is composed by all the remaining particles, $E=P\cup (\cup
_{i=p+1}^{N}P_{i})$. This decomposition results in the system decoherence
when the number $p$ is very large.\bigskip

We have also studied a generalization of the spin-bath model, where a whole
closed system was split into an open many-spin system and its environment.%
\cite{JPA2} In this case we studied different partitions of the whole system
and identified those for which the selected system does not decohere. As
stressed in that work, this might help us to define clusters of particles
that can be used to store q-bits.

The results obtained in both cases allowed us to argue that Zurek's
\textquotedblleft looming big\textquotedblright\ problem is actually a
pseudo-problem, which is simply dissolved by the fact that the split of a
closed quantum system into an open subsystem and its environment is just a
way of selecting a particular space of relevant observables of the whole
closed system. But since there are many different spaces of relevant
observables depending on the observational viewpoint adopted, the same
closed system can be decomposed in many different ways: each decomposition
represents a decision about which degrees of freedom are relevant and which
can be disregarded in each case. And since there is no privileged or
\textquotedblleft essential\textquotedblright\ decomposition, there is no
need of an unequivocal criterion for deciding where to place the cut between
\textquotedblleft the\textquotedblright\ open system and \textquotedblleft
the\textquotedblright\ environment.\ Summing up, decoherence is a phenomenon 
\textit{relative} to the relevant observables selected in each particular
case. The only essential physical fact is that, among all the observational
viewpoints that may be adopted to study a quantum system, some of them
determine subspaces of relevant observables for which the system decoheres.

Another conceptual difficulty of the EID approach relies on its definition
of the pointer basis. This basis is clearly characterized in measurements
situations, where the self-Hamiltonian of the system can be neglected and
the evolution is completely dominated by the interaction Hamiltonian. In
those cases, the pointer basis is given by the eigenstates of the
interaction Hamiltonian.\cite{Zurek-1981} However, there are two further
regimes, differing in the relative strength of the system's self-Hamiltonian
and the interaction Hamiltonian, where the pointer basis lacks a general
definition.\cite{Paz-Zurek 1999} Our present research is directed to the
search of a general and precise definition of the pointer basis of
decoherence.

\section{Modal-Hamiltonian interpretation of quantum mechanics}

Our work on decoherence from a closed-system perspective taught us that the
decomposition of the total Hamiltonian has to be studied in detail in each
case, in order to know whether the system of interest resulting from the
partition decoheres or not under the action of its self-Hamiltonian and the
interaction Hamiltonian. Once we acknowledged the central role played by the
Hamiltonian in decoherence, the natural further step was to ask ourselves
whether it plays the same central role in interpretation. This question led
us to formulate our modal-Hamitonian interpretation (MHI) of QM,\cite{LC}$%
^{-}$\cite{LASC} which belongs to the modal family:\cite{libro modal} it is
a realist, non-collapse interpretation, according to which the quantum state
describes the possible properties of a system but not its actual properties.
Here we will only recall its main interpretative postulates.

The first step is to identify the systems that populate the quantum world.
By adopting an algebraic perspective, a quantum system is defined as:

\textbf{Systems postulate (SP):} \textit{A quantum system }$\mathcal{S}$%
\textit{\ is represented by a pair }$(\mathcal{O},\,H)$\textit{\ such that
(i) }$\mathcal{O}$\textit{\ is a space of self-adjoint operators on a
Hilbert space }$\mathcal{H}$\textit{, representing the observables of the
system, (ii) }$H\in \mathcal{O}$\textit{\ is the time-independent
Hamiltonian of the system }$\mathcal{S}$\textit{, and (iii) if }$\rho
_{0}\in \mathcal{O}^{\prime }$\textit{\ (where }$\mathcal{O}^{\prime }$%
\textit{\ is the dual space of }$\mathcal{O}$\textit{) is the initial state
of }$\mathcal{S}$\textit{, it evolves according to the Schr\"{o}dinger
equation in its von Neumann version.} 

Of course, any quantum system can be partitioned in many ways; however, not
any partition will lead to parts which are, in turn, quantum systems.\cite%
{Harshman-1}$^{,}$\cite{Harshman-2} On this basis, a composite system is
defined as:

\textbf{Composite systems postulate (CSP):}\textit{\ A quantum system
represented by }$\mathcal{S}:\;(\mathcal{O}\,,\,H)$\textit{, with initial
state }$\rho _{0}\in \mathcal{O}^{\prime }$\textit{, is composite when it
can be partitioned into two quantum systems }$\mathcal{S}^{1}:\;(\mathcal{O}%
^{1},\,H^{1})$\textit{\ and }$\mathcal{S}^{2}:\;(\mathcal{O}^{2}\,,\,H^{2})$%
\textit{\ such that (i) }$\mathcal{O}=\mathcal{O}^{1}\otimes \mathcal{O}^{2}$%
\textit{, and (ii) }$H=H^{1}\otimes I^{2}+I^{1}\otimes H^{2}$\textit{,
(where }$I^{1}$\textit{\ and }$I^{2}$\textit{\ are the identity operators in
the corresponding tensor product spaces). In this case, the initial states
of }$\mathcal{S}^{1}$\textit{\ and }$\mathcal{S}^{2}$\textit{\ are obtained
as the partial traces }$\rho _{0}^{1}=Tr_{\mathrm{(}2\mathrm{)}}\rho _{0}$%
\textit{\ and }$\rho _{0}^{2}=Tr_{\mathrm{(}1\mathrm{)}}\rho _{0}$\textit{;
we say that }$\mathcal{S}^{1}$\textit{\ and }$\mathcal{S}^{2}$\textit{\ are
subsystems of the composite system, }$\mathcal{S}=\mathcal{S}^{1}\cup 
\mathcal{S}^{2}$\textit{. If the system is not composite, it is elemental. } 

Since the contextuality of QM, as implied by the Kochen-Specker theorem,\cite%
{K-S} prevents us from consistently assigning actual values to all the
observables of a quantum system in a given state, the second step is to
identify the \textit{preferred context}, that is, the set of the
actual-valued observables of the system. Whereas the different rules of
actual-value ascription proposed by previous modal interpretations rely on
mathematical properties of the theory, our MHI places an element with a
clear physical meaning, the Hamiltonian, at the heart of its rule:

\textbf{Actualization rule (AR):}\textit{\ Given an elemental quantum system
represented by }$\mathcal{S}:\;(\mathcal{O}\,,\,H)$\textit{, the
actual-valued observables of }$\mathcal{S}$\textit{\ are }$H$\textit{\ and
all the observables commuting with }$H$\textit{\ and having, at least, the
same symmetries as }$H$\textit{. }

This preferred context where actualization occurs is independent of time:
the actual-valued observables always commute with the Hamiltonian and,
therefore, they are constants of motion of the system. In other words, the
observables that receive actual values are the same during all the
\textquotedblleft life\textquotedblright\ of the quantum system as such $-$%
precisely, as a closed system$-$: there is no need of accounting for the
dynamics of the actual properties of the quantum system as in other modal
interpretations.\cite{Vermaas}

The fact that the Hamiltonian always belongs to the preferred context agrees
with the many physical cases where the energy has definite value. The MHI
has been applied to several well-known physical situations (hydrogen atom,
Zeeman effect, fine structure, etc.), leading to results consistent with
experimental evidence.\cite{LC} Moreover, it has proved to be effective for
solving the measurement problem, both in its ideal and its non-ideal
versions,\cite{LC} solving the deep challenges that non-ideal measurements
pose to other modal interpretations.\cite{Elby}$^{,}$\cite{Albert-Loewer} In
particular, the MHI distinguishes between reliable and non-reliable
non-ideal measurements.\cite{LC} Furthermore, in spite of the fact that MHI
applies to closed systems, we have proved its compatibility with EID.\cite%
{Manuscrito}$^{,}$\cite{PoS}

Once the MHI was clearly formulated, our further question was whether it
satisfies the Galilean invariance of the theory. In fact, any continuous
transformation admits two interpretations. Under the active interpretation,
the transformation corresponds to a change from one system to another $-$%
transformed$-$ system; under the passive interpretation, the transformation
consists in a change of the viewpoint $-$reference frame$-$ from which the
system is described.\cite{Brading} Nevertheless, in both cases the validity
of a group of symmetry transformations expresses the fact that the identity
and the behavior of the system are not altered by the application of the
transformations: in the active interpretation language, the original and the
transformed systems are equivalent; in the passive interpretation language,
the original and the transformed reference frames are equivalent. Then, any
realist interpretation should agree with that physical fact: the rule of
actual-value ascription should select a set of actual-valued observables
that remains unaltered under the transformations. Since the Casimir
operators of the central-extended Galilei group are invariant under all the
transformations of the group, one can reasonably expect that those Casimir
operators belong to the preferred context.

As we have seen, the preferred context selected by AR only depends on the
Hamiltonian of the system. Then, the requirement of invariance of the
preferred context under the Galilei transformations is directly fulfilled
when the Hamiltonian is invariant, that is, in the case of
time-displacement, space-displacement and space-rotation:%
\begin{eqnarray}
H^{\prime } &=&e^{iH\tau }H\,e^{-iH\tau }=H\ \text{\ (since }\left[ H,H%
\right] =0\text{)}  \label{3-1} \\
H^{\prime } &=&e^{iP_{i}r_{i}}H\,e^{-iP_{i}r_{i}}=H\ \ \text{(since }\left[
P_{i},H\right] =0\text{) }  \label{3-2} \\
H^{\prime } &=&e^{iJ_{i}\theta _{i}}H\,e^{-iJ_{i}\theta _{i}}=H\ \ \text{%
(since }\left[ J_{i},H\right] =0\text{)}  \label{3-3}
\end{eqnarray}%
However, it is not clear that the requirement completely holds, since the
Hamiltonian is not invariant under Galilei-boosts. In fact, under a
Galilei-boost corresponding to a velocity $u_{x}$, $H$ changes as%
\begin{equation}
H^{\prime }=e^{iK_{x}^{(G)}u_{x}}H\,e^{-iK_{x}^{(G)}u_{x}}\neq H\ \ \text{%
(since }\left[ K_{x}^{(G)},H\right] =iP_{x}\neq 0\text{)}  \label{3-4}
\end{equation}%
Nevertheless, when space is homogeneous and isotropic, a Galilei-boost only
introduces a change in the subsystem that carries the kinetic energy of
translation: the internal energy $W$ remains unaltered under the
transformation. This should not sound surprising to the extent that $W$ $-$%
multiplied by the scalar mass $m-$ is a Casimir operator of the
central-extended Galilei group. On this basis, we can reformulate AR in an
explicit Galilei-invariant form in terms of the Casimir operators of the
central-extended group:

\textbf{Actualization rule'\ (AR')}\textit{: Given a quantum system free
from external fields and represented by }$\mathcal{S}:\;(\mathcal{O}\,,\,H)$%
\textit{, its actual-valued observables are the observables }$C_{i}^{G}$%
\textit{\ represented by the Casimir operators of the central-extended
Galilei group in the corresponding irreducible representation, and all the
observables commuting with the }$C_{i}^{G}$\textit{\ and having, at least,
the same symmetries as the }$C_{i}^{G}$\textit{. }

Since the observables\textit{\ }$C_{i}^{G}$\textit{\ } $-$in the reference
frame of the center of mass$-$ are $M$, $mW$ and $m^{2}S^{2}$, this new
version AR' is in agreement with the original AR when applied to a system
free from external fields:\cite{ACL}$^{-}$\cite{LCA}

\begin{itemize}
\item The actual-valuedness of $M$ and $S^{2}$, postulated by AR', follows
from AR: these observables commute with $H$ and do not break its symmetries
because, in non-relativistic QM, both are multiples of the identity in any
irreducible representation.

\item The actual-valuedness of $W$ might seem to be in conflict with AR
because $W$ is not the Hamiltonian: whereas $W$ is Galilei-invariant, $H$
changes under the action of a Galilei-boost. However, this is not a real
obstacle because a Galilei-boost transformation only introduces a change in
the subsystem that carries the kinetic energy of translation, which can be
considered a mere shift in an energy defined up to a constant.\cite{ACL}$%
^{,} $\cite{LCA}
\end{itemize}

Summing up, the application of AR' leads to reasonable results, since the
actual-valued observables turn out to be invariant and, therefore, objective
magnitudes. The assumption of a strong link between invariance and
objectivity is rooted in a natural idea: what is objective should not depend
on the particular perspective used for the description; or, in
group-theoretical terms, what is objective according to a theory is what is
invariant under the symmetry group of the theory. This idea is not new: it
was widely discussed in the context of special and general relativity with
respect to the ontological status of space and time,\cite{Minkowski} and
since then it reappeared in several works. \cite{Weyl}$^{-}$\cite{Earman-2}
>From this perspective, AR says that the observables that acquire actual
values are those representing objective magnitudes. On the other hand, from
any realist viewpoint, the fact that certain observables acquire an actual
value is an objective fact in the behavior of the system; therefore, the set
of actual-valued observables selected by a realist interpretation must be
also Galilean-invariant. But the Galilean-invariant observables are always
functions of the Casimir operators of the Galilean group. As a consequence,
one is led to the conclusion that any realist interpretation that intends to
preserve the objectivity of actualization may not stand very far from the
modal-Hamiltonian interpretation.

When AR is expressed in simple group terms, one can expect that it can be
extrapolated to any quantum theory endowed with a symmetry group. In
particular, the actual-valued observables of a system in quantum field
theory would be those represented by the Casimir operators of the Poincar%
\'{e} group and of the internal symmetry group. On this basis, in a recent
paper we presented an alternative version of the non-relativistic limit of
the centrally extended Poincar\'{e} group and its consequences for
interpretive problems.\cite{ACR}

As it is well known, the Galilei group can be recovered from the Poincar\'{e}
group by means of In\"{o}n\"{u}-Wigner contraction.\cite{LL} It is therefore
natural to ask whether such a situation can be generalized to the
central-extended Galilei group, which is the relevant group in QM. However,
since the Poincar\'{e} group does not admit nontrivial central extensions,%
\cite{cari} we have to define a generalized In\"{o}n\"{u}-Wigner contraction
from a trivial extension of the Poincar\'{e} group whose generators are $H$, 
$P_{i}$, $J_{i}$ and $K_{P_{i}}$ (where the last ones are the Lorentz
boosts). With this purpose, we extend the group trivially, i.e., in such a
way that all the generators of Poincar\'{e} group commute with a trivial
central charge $M$. The basis of the resulting new algebra $%
I^{M}SO(1,3)=ISO(1,3)\times \left\langle M\right\rangle $ is $\left\{ H\text{%
, }P_{i}\text{, }J_{i}\text{, }K_{P_{i}}\text{, }M\right\} $. Then, we
perform the following change of the generators basis $\overline{H}=H-M$. In
the new basis $\left\{ \overline{H}\text{, }P_{i}\text{, }J_{i}\text{, }%
K_{P_{i}}\text{, }M\right\} $ all the commutators of the Poincar\'{e} group
remain the same, with the only exception of%
\begin{equation}
\left[ P_{i},K_{P_{j}}\right] =-i\delta _{ij}H=-i\delta _{ij}(\overline{H}+M)
\label{3-5}
\end{equation}%
The contraction is determined by the rescaling transformations (in the basis 
$\left\{ \overline{H}\text{, }P_{i}\text{, }J_{i}\text{, }K_{P_{i}}\text{, }%
M\right\} $) defined by%
\begin{equation}
J_{i}^{\prime }=J_{i}\text{, \ \ }P_{i}^{\prime }=\varepsilon P_{i}\text{, \
\ }K_{P_{i}}^{\prime }=\varepsilon K_{P_{i}}\text{, \ \ }\overline{H}%
^{\prime }=\overline{H}\text{, \ \ \ }M^{\prime }=\varepsilon ^{2}M\text{ \
\ \ }  \label{3-6}
\end{equation}%
The space isotropy remains unchanged by this rescaling transformation and%
\begin{equation}
\left[ P_{i}^{\prime },K_{P_{j}}^{\prime }\right] =-i\delta
_{ij}(\varepsilon ^{2}\overline{H}^{\prime }+M^{\prime })\ \ \ \ \ so\ \ \ \
\ \ \ \ \underset{\varepsilon \rightarrow 0}{\lim }\left[ P_{i}^{\prime
},K_{P_{j}}^{\prime }\right] =-i\delta _{ij}M^{\prime }  \label{3-7}
\end{equation}%
Therefore, it turns out to be clear that the contracted algebra is
isomorphic to the extension of the Galilei algebra. On the basis of this
result, we have also proved that the Casimir operators of the trivially
extended Poincar\'{e} group contract naturally to the Casimir operators of
the extended Galilei group.\cite{ACR}

Summing up, when AR is expressed in its explicit Galilei-invariant form AR',
it leads to a physically reasonable result: the actual-valued observables
are those represented by the Casimir operators of the mass central-extended
Galilei group. The natural strategy is to extrapolate the interpretation to
the relativistic realm by replacing the Galilei group with the Poincar\'{e}
group. But when one takes into account that the relevant group of
non-relativistic QM is not the Galilei group but its central extension, the
mere replacement of the relevant group is not sufficient: one has to show
also that the actual-valued observables in the relativistic and the
non-relativistic cases are related through the adequate limit. As a
consequence, the Poincar\'{e} group has to be trivially extended, in order
to show that the limit between the corresponding Casimir operators holds,
and this result counts in favor of the proposed extrapolation of our MHI to
non-relativistic QM. Furthermore, this result is physically reasonable
because mass and spin are properties supposed to be always possessed by any
elemental particle,\cite{Haag} and they are two of the properties that
contribute to the classification of elemental particles. At present we are
working on a further extrapolation of the MHI to the standard model.

\section{The ontological picture of the quantum world}

In general, the discussions about modal interpretations are concerned with
the traditional problems, as the measurement problem and the no-go theorems.
But these are not the only relevant issues: one should not forget the
ontological question about the structure of the world referred to by QM.

All modal interpretations rely on a common assumption: QM does not describe
what is the case, but rather what may be the case. The problem of the nature
of possibility is as old as philosophy itself. Since Aristotle's time to
nowadays, however, two general conceptions can be identified. On the one
hand, \textit{actualism} reduces possibility to actuality. This was the
position of Diodorus Cronus, who defined \textquotedblleft \textit{the
possible as that which either is or will be}\textquotedblright . \cite{K-K}\
This view survived up to 20$^{th}$ century; for instance, for Russell
`possible'\ means `sometimes', whereas `necessary'\ means `always'.\cite%
{Russell} On the other hand, \textit{possibilism} conceives possibility as
an ontologically irreducible feature of reality. From this perspective, the
stoic Crissipus defined possible as \textquotedblleft \textit{that which is
not prevented by anything from happening even if it does not happen}%
\textquotedblright .\cite{Bunge} In present day metaphysics, the debate
actualism-possibilism is still alive. For the actualists, the adjective
`actual'\ is redundant: non-actual possible items (objects, properties,
facts, etc.) do not exist. According to the possibilists, on the contrary,
not every possible item is an actual item: possible items---\textit{%
possibilia}---constitute a basic ontological category.\cite{Menzel}

For our MHI, probabilities measure ontological propensities, which embody a
possibilist, non-actualist possibility: a possible fact does not need to
become actual to be real. This possibility is defined by the postulates of
QM and is not reducible to actuality. This means that reality spreads out in
two realms, the realm of possibility and the realm of actuality. In
Aristotelian terms, being can be said in different ways: as possible being
or as actual being, and none of them is reducible to the other. Moreover,
the \textit{ontological structure of the realm of possibility} is embodied
in the definition of the elemental quantum system $\mathcal{S}:\;(\mathcal{O}%
\,,\,H)$, with its initial state $\rho _{0}$: (i) the space of observables $%
\mathcal{O}\,$ identifies all the \textit{possible type-properties}
(observables) with their corresponding \textit{possible case-properties}
(eigenvalues), and (ii) the initial state $\rho _{0}$ codifies the measures
of the \textit{propensities to actualization} of all the possible
case-properties at the initial time, propensities that evolve
deterministically according to the Schr\"{o}dinger equation.

The fact that propensities belong to the realm of possibility does not mean
that they do not have physical consequences in the realm of actuality. On
the contrary, propensities produce definite effects on actual reality even
if they never become actual. An interesting manifestation of such
effectiveness is the case of the so-called \textquotedblleft \textit{%
non-interacting experiments}\textquotedblright ,\cite{E-V}$^{,}$\cite%
{Vaidman} where non-actualized possibilities can be used in practice, for
instance, to test bombs without exploding them.\cite{Penrose} This shows
that possibility is a way in which reality manifests itself, a way
independent of and not less real than actuality.

One of the main areas of controversy in contemporary metaphysics is the
problem of the nature of individual objects: is an individual a substratum
supporting properties or a mere \textquotedblleft bundle\textquotedblright\
of properties?\cite{Loux} The idea of a substratum acting as a bearer of
properties has pervaded the history of philosophy: it is present under
different forms in Aristotle's \textquotedblleft primary
substance\textquotedblright , in Locke's doctrine of \textquotedblleft
substance in general\textquotedblright\ or in Leibniz's monads.
Nevertheless, many philosophers belonging to the empiricist tradition, from
Hume to Russell, Ayer and Goodman, have considered the posit of a
characterless substratum as a metaphysical abuse. As a consequence, they
adopted some version of the \textit{bundle theory}, according to which an
individual is nothing but a bundle of properties: properties have
metaphysical priority over individuals and, therefore, they are the
fundamental items of the ontology.

In the Hilbert space formalism, states have logical priority over
observables since observables apply to states. This logical priority favors
the picture of an ontology of substances and properties, with the
traditional priority of substances over properties. Our MHI, on the
contrary, is based on the algebraic formalism, where the basic elements are
observables and states are functionals over the space of observables. Then,
the MHI favors the bundle theory, that is, an ontology of properties, where
the category of substance is absent.

According to the traditional versions of the bundle theory, an individual is
the convergence of certain case-properties, under the assumption that all
the type-properties are determined in the actual realm. For instance, a
particular billiard ball is the convergence of a definite value of position,
a definite shape, say round, a definite color, say white, etc. Then, the
properties taken into account are always actual properties: bundle theories
identify individuals with bundles of actual properties. In QM, on the
contrary, the Kochen-Specker theorem prevents the assignment of
case-properties (eigenvalues) to all the type-properties (observables) of
the system in a non-contradictory manner. Therefore, the classical idea of a
bundle of actual properties does not work for the quantum ontology.

If, from the perspective of the MHI, the quantum world unfolds into two
irreducible realms, the realm of possibility has to be taken into account
when deciding what kind of properties constitutes the quantum bundle. Since
the quantum system is identified by its space of observables, which
represent possible properties, an individual quantum system turns out to be 
\textit{a bundle of possible properties}: it inhabits the realm of
possibility, which is as real as the realm of actuality.

This interpretation of quantum individual systems has the advantage of being
immune to the challenge represented by the Kochen-Specker theorem, since
this theorem imposes no restriction on possibilities. Moreover, it seems
reasonable to expect that this conception of individual supplies the basis
for solving the problem of the indistinguishability of \textquotedblleft
identical particles\textquotedblright ,\cite{French-Krause} introduced in
the formalism as an \textit{ad hoc} restriction on the set of states. At
present, we are working on this problem: if the traditional assumption of
substantial objects, which preserve their individuality when considered in
collections, is the main obstacle to explain quantum statistics, the
conception of the quantum system as a bundle of possible properties seems to
offer a promising starting point in the search for a solution of the problem.

Summing up, from our interpretational perspective, the talk of individual
entities as electrons or photons and their interactions can be retained only
in a metaphorical sense. In fact, in the quantum framework even the number
of particles is represented by an observable $N$, which is subject to the
same theoretical constraints as any other observable of the system; this
leads, specially in quantum field theory, to the possibility of states that
are superpositions of different particle numbers.\cite{Butterfield}
Therefore, the number of particles $N$ has an actual definite value only in
some cases, but it is indefinite in others. This fact, puzzling from an
ontology populated by substantial objects, is deprived of mystery when
viewed from our ontological perspective. The quantum system is not a
substantial individual, but a bundle of possible properties. The particle
picture, with a definite number of particles, is only a contextual picture
valid exclusively when the observable $N$ is picked out by the preferred
context. In this case, we could metaphorically retain the idea of a
composite system composed of individual particles that interact to each
other. But in the remaining cases, this idea proves to be completely
inadequate, even in a metaphorical sense.

\section{Concluding remarks}

We hope that this journey through the main contributions of Castagnino in
the field of the foundations of QM supplies an idea of the active work that
he and his research group are developing. Nevertheless, we do not want to
finish this review without recalling the rest of the areas of the philosophy
of physics where he has fruitfully produced: time's arrow,\cite{AoT-1}$^{-}$%
\cite{AoT-8} time-asymmetric QM,\cite{TAQM-1}$^{,}$\cite{TAQM-2} quantum
chaos,\cite{QC-1}$^{,}$\cite{QC-2} and even philosophy of chemistry.\cite%
{Chem} Not bad for a baby philosopher. However, this is not surprising when
coming from an even-eager spirit as Mario Castagnino.


\begin{thebibliography}{99}
\bibitem{CL-1} M. Castagnino and R. Laura, \textit{Phys. Rev}.\textit{\ A}, 
\textbf{56,} 108, 1997.

\bibitem{CL-2} R. Laura and M. Castagnino, \textit{Phys. Rev.} \textit{A}, 
\textbf{57}, 4140, 1998.

\bibitem{CL-3} R. Laura and M. Castagnino, \textit{Phys. Rev. E}, \textbf{57}%
, 3948, 1998.

\bibitem{CL-4} M. Castagnino, \textit{Int. Jour. Theor. Phys.}, \textbf{38},
1333, 1999.

\bibitem{CL-5} M. Castagnino and R. Laura, \textit{Phys. Rev. A}, \textbf{62}%
, 022107, 2000.

\bibitem{CL-6} M. Castagnino and R. Laura, \textit{Int. Jour. Theor. Phys.}, 
\textbf{39}, 1767, 2000.

\bibitem{SID-1} M. Castagnino and O. Lombardi, \textit{Int. Jour. Theor.
Phys.}, \textbf{42}, 1281, 2003.

\bibitem{SID-2} M. Castagnino and O. Lombardi, \textit{Stud. Hist. Phil.
Mod. Phys}, \textbf{35}, 73, 2004.

\bibitem{SID-3} M. Castagnino, \textit{Physica A}, \textbf{335}, 511, 2004.

\bibitem{SID-4} M. Castagnino and A. Ordo\~{n}ez, \textit{Int. Jour. Theor.
Phys.}, \textbf{43}, 695, 2004.

\bibitem{SID-5} M. Castagnino, \textit{Braz. Jour. Phys}., \textbf{35}, 375,
2005.

\bibitem{SID-6} M. Castagnino and O. Lombardi, \textit{Phys. Rev. A}, 
\textbf{72}, \# 012102, 2005.

\bibitem{SID-7} M. Castagnino and O. Lombardi, \textit{Phil. Scie}., \textbf{%
72}, 764, 2005.

\bibitem{SID-8} M. Castagnino, \textit{Phys. Lett. A}, \textbf{357}, 97,
2006.

\bibitem{SID-9} M. Castagnino and M. Gadella, \textit{Found. Phys}., \textbf{%
36}, 920, 2006.

\bibitem{SID-10} M. Castagnino and O. Lombardi, \textit{Physica A}, 388,
247, 2009.

\bibitem{Paz-Zurek} J. P. Paz and W. H. Zurek, \textquotedblleft
Environment-induced decoherence and the transition from quantum to
classical\textquotedblright , in Dieter Heiss (ed.), \textit{Lecture Notes
in Physics, Vol. 587}, Heidelberg-Berlin: Springer, 2002.

\bibitem{Zurek-2003} W. H. Zurek, \textit{Rev. Mod. Phys}., \textbf{75},
715, 2003.

\bibitem{GTFD-1} M. Castagnino, R. Laura and O. Lombardi, \textit{Phil. Scie}%
., \textbf{74}, 968, 2007.

\bibitem{GTFD-2} M. Castagnino, S. Fortin, R. Laura and O. Lombardi, \textit{%
Class. Quant. Grav.}, \textbf{25}, 154002, 2008.

\bibitem{MPLA-random} M. Castagnino, S. Fortin amd O. Lombardi, \textit{Mod.
Phys. Lett. A}, \textbf{25}, 611, 2010

\bibitem{Omnes-2001} R. Omn\`{e}s, \textquotedblleft Decoherence: an
irreversible process\textquotedblright , arXiv:quant-ph/0106006, 2001.

\bibitem{Omnes-2002} R. Omn\`{e}s, \textit{Phys. Rev. A}, \textbf{65},
052119, 2002.

\bibitem{van Kampen} N. G. van Kampen, \textit{Physica}, \textbf{20,} 603,
1954.

\bibitem{Daneri} A. Daneri, A. Loinger and G. Prosperi,\textit{\ Nucl. Phys}%
., \textbf{33, }297, 1962.

\bibitem{Zeh-1} H. D. Zeh, \textit{Found. Phys.}, \textbf{1}, 69, 1970.

\bibitem{Zeh-2} H. D. Zeh, \textquotedblright On the irreversibility of time
and observation in quantum theory\textquotedblright , in B. d'Espagnat
(ed.), \textit{Foundations of Quantum Mechanics}, Academic Press, New York,
1971.

\bibitem{Zeh-3} H. D. Zeh, \textit{Found. Phys.}, \textbf{3}, 109, 1973.

\bibitem{Omnes-1994} R. Omn\`{e}s, \textit{The Interpretation of Quantum
Mechanics}, Princeton: Princeton University Press, 1994.

\bibitem{Omnes-1999} R. Omn\`{e}s, \textit{Understanding Quantum Mechanics},
Princeton: Princeton University Press, 1999.

\bibitem{Zurek-cut} W. H. Zurek, \textit{Phil. Trans. R. Soc.}, \textbf{A356}%
, 1793, 1998.

\bibitem{Zurek-1982} W. H. Zurek, \textit{Phys. Rev. D}, \textbf{26}, 1862,
1982.

\bibitem{MPLA} M. Castagnino, S. Fortin and O. Lombardi, \textit{Mod. Phys.
Lett. A}, \textbf{25, }1431, 2010.

\bibitem{JPA2} M. Castagnino, S. Fortin, and O. Lombardi,\textit{\ J. Phys.
A: Math. Theor.}, \textbf{43}, 065304, 2010.

\bibitem{Zurek-1981} W. H. Zurek, \textit{Phys. Rev. D}, \textbf{24}, 1516,
1981.

\bibitem{Paz-Zurek 1999} J. P. Paz and W. H. Zurek, \textit{Phys. Rev. Lett}%
., \textbf{82}, 5181, 1999.

\bibitem{LC} O. Lombardi and M. Castagnino, \textit{Stud. Hist. Phil. Mod.
Phys}., \textbf{39}, 380, 2008 .

\bibitem{CL} M. Castagnino and O. Lombardi, \textit{J. Phys. Conf. Ser.}, 
\textbf{28}, 012014, 2008.

\bibitem{LASC} O. Lombardi, S. Fortin, J. S. Ardenghi and M. Castagnino, 
\textit{Introduction to the Modal-Hamiltonian Interpretation of Quantum
Mechanics},\ New York: Nova Science, forthcoming.

\bibitem{libro modal} D. Dieks and P. E. Vermaas, \textit{The Modal
Interpretation of Quantum Mechanics}, Dordrecht: Kluwer Academic Publishers,
1998.

\bibitem{Harshman-1} N. L. Harshman and S. Wickramasekara, \textit{Phys.
Rev. Lett.}, \textbf{98}, 080406, 2007.

\bibitem{Harshman-2} N. L. Harshman and S. Wickramasekara, \textit{Open
Systems and Information Dynamics}, \textbf{14}, 341, 2007.

\bibitem{K-S} S. Kochen and E. Specker, \textit{Jour. Math. Mech}., \textbf{%
17}, 59, 1967.

\bibitem{Vermaas} P. E. Vermaas, \textit{Stud. Hist. Phil. Mod. Phys}., 
\textbf{27}, 133, 1996.

\bibitem{Elby} A. Elby, \textit{Found. Phys. Lett}., \textbf{6}, 5, 1993.

\bibitem{Albert-Loewer} D. Albert and B. Loewer, \textit{Found. Phys. Lett}, 
\textbf{6}, 297, 1993.

\bibitem{Manuscrito} O. Lombardi, \textquotedblleft The central role of the
Hamiltonian in quantum mechanics: decoherence and
interpretation\textquotedblright , \textit{Manuscrito}, forthcoming.

\bibitem{PoS} O. Lombardi, S. Fortin, M. Castagnino and J. S. Ardenghi,
\textquotedblleft Compatibility between environment-induced decoherence and
the modal-Hamiltonian interpretation of quantum mechanics\textquotedblright
, \textit{Phil. Scie}., forthcoming.

\bibitem{Brading} K. Brading and E. Castellani, \textquotedblleft Symmetries
and invariances in classical physics\textquotedblright , in J. Butterfield
\& J. Earman (eds.), \textit{Philosophy of Physics}, Amsterdam:
North-Holland, 2007.

\bibitem{ACL} J. S. Ardenghi, M. Castagnino and O. Lombardi, \textit{Found.
Phys.}, \textbf{39}, 1023, 2009.

\bibitem{Theoria} O. Lombardi, M. Castagnino and J. S. Ardenghi, \textit{%
Theoria}, \textbf{24}, 5, 2009.

\bibitem{LCA} O. Lombardi, M. Castagnino and J. S. Ardenghi, \textit{Stud.
Hist. Phil. Mod. Phys}., \textbf{41}, 93, 2010.

\bibitem{Minkowski} H. Minkowski, \textquotedblleft Space and
time\textquotedblright , in W. Perrett and G. B. Jeffrey (eds.),\textit{\
The Principle of Relativity. A Collection of Original Memoirs on the Special
and General Theory of Relativity,} New York: Dover, 1923.

\bibitem{Weyl} H. Weyl, \textit{Symmetry}, Princeton: Princeton University
Press, 1952.

\bibitem{Auyang} S. Y. Auyang, \textit{How is Quantum Field Theory Possible?,%
} Oxford: Oxford University Press, 1995.

\bibitem{Nozick} R. Nozick, \textit{Invariances: The Structure of the
Objective World}, Harvard: Harvard University Press, 2001.

\bibitem{Brading-Castellani} K. Brading \& E. Castellani (eds.), \textit{%
Symmetries in Physics: Philosophical Reflections}, Cambridge: Cambridge
University Press, 2003.

\bibitem{Earman-1} J. Earman, \textit{Phil. Scie}., \textbf{69}, S209, 2002.

\bibitem{Earman-2} J. Earman, \textit{Phil. Scie}., \textbf{71}, 1227, 2004.

\bibitem{ACR} J. S. Ardenghi, M. Castagnino and R. Campoamor-Stursberg, 
\textit{Jour. Math. Phys.}, \textbf{50}, 103526, 2009.

\bibitem{LL} H. Bacry and J.-M. Levy L\'{e}blond, \textit{Journ. Math. Phys}%
., \textbf{9}, 1605, 1968.

\bibitem{cari} J. F. Cari\~{n}ena, M. A. del Olmo and M. Santander,\textit{\
Jour. Phys. A}, \textbf{14}, 1, 1981.

\bibitem{Haag} R. Haag, \textquotedblleft Questions in quantum physics: a
personal view\textquotedblright , arXiv:hep-th/0001006, 2006.

\bibitem{K-K} W. Kneale and M. Kneale, \textit{The Development of Logic},
Oxford: Clarendon Press, 1962.

\bibitem{Russell} B. Russell, \textit{Introduction to Mathematical Philosophy%
}, London: Allen and Unwin, 1919.

\bibitem{Bunge} M. Bunge, \textit{Treatise on Basic Philosophy, Vol.3:
Ontology I}, Dordrecht: Reidel, 1977.

\bibitem{Menzel} C. Menzel, \textquotedblleft Actualism\textquotedblright ,
in E. N. Zalta (ed.), \textit{The Stanford Encyclopedia of Philosophy}
(Spring 2007 Edition), http://plato.stanford.edu, 2007.

\bibitem{E-V} A. C. Elitzur and L. Vaidman, \textit{Found. Phys.}, \textbf{23%
}, 987, 1993.

\bibitem{Vaidman} L. Vaidman, \textquotedblleft On the paradoxical aspects
of new quantum experiments\textquotedblright , \textit{Proceedings of the
1994 Biennial Meeting of the Philosophy of Science Association, Vol. 1},
East Lansing: Philosophy of Science Association, 1994.

\bibitem{Penrose} R. Penrose, \textit{Shadows of the Mind}, Oxford: Oxford
University Press, 1994.

\bibitem{Loux} M. Loux, \textit{Metaphysics. A Contemporary Introduction},
London: Routledge, 1998.

\bibitem{French-Krause} S. French and D. Krause, \textit{Identity in
Physics: A Historical, Philosophical and Formal Analysis}, Oxford: Oxford
University Press, 2006.

\bibitem{Butterfield} J. Butterfield, \textquotedblleft Interpretation and
identity in quantum theory\textquotedblright , \textit{Stud. Hist. Phil.
Scie.}, \textbf{24}, 443, 1993.

\bibitem{AoT-1} M. Castagnino, L. Lara and O. Lombardi, \textit{Class.
Quant. Grav.}, \textbf{20}, 369, 2003.

\bibitem{AoT-2} M. Castagnino, O. Lombardi and L. Lara, \textit{Found. Phys.}%
, \textbf{33}, 877, 2003.

\bibitem{AoT-3} M. Castagnino, L. Lara and O. Lombardi, \textit{Int. Jour.
Theor. Phys.}, \textbf{42}, 2487, 2003.

\bibitem{AoT-4} M. Castagnino and O. Lombardi, \textit{Jour. Physics A}, 
\textbf{37}, 4445, 2004.

\bibitem{AoT-5} M. Castagnino and O. Lombardi, \textquotedblleft Time
asymmetry as universe asymmetry\textquotedblright , in O. Descalzi, J. Mart%
\'{\i}nez and S. Rica (eds.), \textit{Instabilities and Nonequilibrium
Structures IX}, Dordrecht: Kluwer, 2004.

\bibitem{AoT-6} M. Castagnino and O. Lombardi, \textquotedblleft A global
and non entropic approach to the problem of the arrow of
time\textquotedblright , in A. Reimer (ed.), \textit{Spacetime Physics
Research Trends. Horizons in World Physics}, New York: Nova Science, 2005.

\bibitem{AoT-7} M. Aiello, M. Castagnino and O. Lombardi, \textit{Found.
Phys.}, \textbf{38}, 257, 2008.

\bibitem{AoT-8} M. Castagnino and O. Lombardi, \textit{Synthese}, \textbf{169%
}, 1, 2009.

\bibitem{TAQM-1} M. Castagnino, M. Gadella and O. Lombardi, \textit{Int.
Stud. Phil. Scie}., \textbf{19}, 223, 2005.

\bibitem{TAQM-2} M. Castagnino, M. Gadella and O. Lombardi, \textit{Found.
Phys.}, \textbf{36}, 407, 2006.

\bibitem{QC-1} M. Castagnino and O. Lombardi, \textit{Chaos, Solitons, and
Fractals}, \textbf{28}, 879, 2006.

\bibitem{QC-2} M. Castagnino and O. Lombardi, \textit{Stud. Hist. Phil. Mod.
Phys.}, \textbf{38}, 482, 2007.

\bibitem{QC-3} M. Castagnino and O. Lombardi, \textit{Physica A}, \textbf{388%
}, 247, 2009.

\bibitem{Chem} O. Lombardi and M. Castagnino, \textquotedblleft Matters are
not so clear on the physical side\textquotedblright , \textit{Found. Chem}.,
forthcoming.
\end{thebibliography}
\end{document}